\documentclass[
               twocolumn,
               noshowpacs,          
               nopreprintnumbers,     
               aps,                 
               prd,                 
               a4paper,             
               superscriptaddress,  
               nofootinbib,         
               tightenlines,        
               floats,floatfix      
               ]{revtex4}

\usepackage{amsmath, amsfonts, amsthm, amssymb, graphicx}
\usepackage{epstopdf}
 \usepackage{slashed}
 \usepackage{graphicx,epsfig}
\usepackage{amsmath}
\usepackage {amssymb}
\usepackage[utf8]{inputenc}
\usepackage{hyperref}
\usepackage[francais]{babel}
\usepackage[T1]{fontenc}

\newcommand{\be}{\begin{eqnarray}}
\newcommand{\ee}{\end{eqnarray}}
\newcommand{\bea}{\begin{eqnarray}}

\newcommand{\eea}{\end{eqnarray}}

\def\b{\theta(\phi)}
\def\gff{\gamma(\phi)}
\def\db{\theta'(\phi)}
\def\dgff{\gamma'(\phi)}

\begin{document}

\title{Supertranslations and Superrotations at the Black Hole Horizon}

\author{Laura Donnay}
\email{laura.donnay-at-ulb.ac.be}
\affiliation{Physique Th\'eorique et Math\'ematique, Universit\'e Libre de Bruxelles and International Solvay Institutes,
Campus Plaine C.P. 231, B-1050 Bruxelles, Belgium.}

\author{Gaston Giribet}
\email{gaston-at-df.uba.ar}
\affiliation{Departamento de F\'isica, Universidad de Buenos Aires FCEN-UBA and IFIBA-CONICET, Ciudad Universitaria, Pabell\'on 1, 1428 Buenos Aires, Argentina.}
\affiliation{Instituto de F\'isica, Pontificia Universidad Cat\'olica de Valpara\'iso, Casilla 4950, Valpara\'iso 05101, Chile.}

\author{Hern\'an A. Gonz\'alez}
\email{hgonzale-at-ulb.ac.be}
\affiliation{Physique Th\'eorique et Math\'ematique, Universit\'e Libre de Bruxelles and International Solvay Institutes,
Campus Plaine C.P. 231, B-1050 Bruxelles, Belgium.}

\author{Miguel Pino}
\email{miguel.pino.r-at-usach.cl}
\affiliation{Departamento de F\'isica, Universidad de Santiago de Chile, Casilla 307, Santiago 9170124, Chile.}


\date{\today}

\begin{abstract}
We show that the asymptotic symmetries close to nonextremal black hole horizons are generated by an extension of supertranslations. This group is generated  by a semidirect sum of Virasoro and Abelian currents. The charges associated with the asymptotic Killing symmetries satisfy the same algebra. When considering the special case of a stationary black hole, the zero mode charges correspond to the angular momentum and the entropy at the horizon.  
\end{abstract}

\maketitle

\section{Introduction} 

Infinite-dimensional symmetries play a central role in the holographic description of black holes. The prototypical example is the microscopic derivation of the entropy of asymptotically AdS$_3$ black holes \cite{Strominger} in terms of the Virasoro algebra at infinity \cite{BrownHenneaux}. Virasoro and affine Kac-Moody algebras also appear in the description of non-AdS black holes in three dimensions \cite{WAdS1, GS, HMT, WAdS2, WAdS3} and, in higher dimensions, they govern the physics in the near horizon of rapidly rotating Kerr black holes \cite{KerrCFT}. 

Recently, Hawking, Perry, and Strominger claimed that nonextremal stationary black holes also exhibit infinite-dimensional symmetries in the near horizon region, known as supertranslations \cite{HPS}, and they suggested that this observation could contribute to solving the information paradox for black holes \cite{Hawking}. The symmetry observed in \cite{HPS} is similar to the one that arises in asymptotically flat spacetimes at null infinity \cite{BMS1, BMS2, BMS3}, usually referred to as Bondi-Metzner-Sachs (BMS). The corresponding algebra is an infinite-dimensional extension of the translation part of Poincar\'e.

In the last years, BMS algebra has been reconsidered in relation to flat space holography \cite{otros1, otros2, Barnich1, Barnich2, Barnich3, Barnich4, Barnich6, Hernan1, Hernan2, Hernan3}; that is, the attempt to extend the AdS/CFT holographic correspondence to asymptotically flat spacetimes. In AdS/CFT, a crucial ingredient is the asymptotic isometry group. From the Anti-de Sitter (AdS) bulk point of view it is seen as the symmetries that preserve the form of the geometry close to the boundary region, where the dual conformal field theory (CFT) is located, while from the point of view of the CFT it corresponds to the local conformal group. In flat space holography, the conformal symmetry at the boundary is replaced by the BMS symmetry at the null infinity region. Furthermore, in addition to BMS super-translations, the symmetries at null infinity include super-rotations and central extensions \cite{Barnich1, Barnich2, Barnich3, Barnich4, Barnich6, PH}; see also \cite{otros3}. In presence of black holes, besides the null infinity region, there exists a second codimension 1 null hypersurface near which the geometry is flat: the black hole event horizon. Therefore, a natural question is whether the features associated with holography, such as the enhanced BMS symmetry, also appear in the near horizon geometry of black holes. In this Letter, we will show that for an adequate choice of boundary conditions, the nearby region to the horizon of a stationary black hole exhibits a generalization of supertranslation, including a semidirect sum with superrotations, represented by Virasoro algebra. In this sense, both supertranslations and superrotations arise close to the horizon. However, this particular extension differs from the extended BMS at null infinity \cite{Barnich2, Barnich3}. 

The Letter is organized as follows: In section \ref{sec:3d}, as \emph{entr\'ee}, we consider the three-dimensional case. This allow us to identify the appropriate boundary conditions at the horizon and construct a family of exact solutions satisfying them. This family includes, as a particular case, the Ba\~{n}ados-Teitelboim-Zanelli (BTZ) black holes \cite{BTZ}. We compute the algebra obeyed by the asymptotic Killing vectors and show that they expand supertranslations in semidirect sum with superrotations. The charges associated with such asymptotic symmetries are shown to expand the same algebra and by evaluating them on the BTZ solution, we verify that they correspond to the angular momentum and the entropy of the black hole. We follow the same strategy in section \ref{sec:3}, where we address the four-dimensional case. We demonstrate that the symmetry group generated by these charges  correspond to two copies of Virasoro algebra and two sets of supertranslations. 
The zero mode conserved quantities of Kerr black hole coincide with the entropy and the angular momentum. 
\section{Asymptotic symmetries at the horizon}
\label{sec:2}
We are interested in studying the symmetries preserved by stationary non-extremal black hole metrics close to an event horizon, first in three dimensions and then we move to the four dimensional case.

\subsection{Three-dimensional analysis}
\label{sec:3d}
The near horizon geometry of three-dimensional black holes can be expressed using Gaussian null coordinates 
\begin{equation}\label{ds2}
ds^2= f dv^2 +2 k dv d\rho + 2h dv d\phi+R^2d\phi^2,
\end{equation}
where $v\in \mathbb{R}$ represents the retarded time, $\rho\geq 0$ is the radial distance to the horizon and $\phi $ is the angular coordinate of period $2\pi $. Functions $f$, $k$, $h$, and $R$ are demanded to obey the following fall-off conditions close to $\rho = 0$:
\begin{equation}
\label{theBC}
\begin{split}
f&= -2\kappa \rho + {\mathcal O}(\rho ^2), \\
k&=1+ {\mathcal O}(\rho ^2), \\
h&= \theta(\phi)\rho+{\mathcal O}(\rho^2 ), \\
R^2&=\gamma(\phi)^2+ \lambda(v,\phi) \rho + {\mathcal O}(\rho^2 ),
\end{split}
\end{equation}
where ${\mathcal O}(\rho^2)$ stands for functions of $v$ and $\phi $ that vanish at short $\rho $ equally or faster than $\rho ^2$, consistent with the near horizon approximation. The metric components $g_{\rho \rho}$ and $g_{\rho \phi}$, which do not appear in \eqref{ds2}, are supposed to be $\mathcal{O}(\rho^2)$. One can verify that this asymptotic behavior is preserved by the asymptotic diffeomorphisms we will consider, see (\ref{killing}) below. In particular, no order $\mathcal{O}(\rho)$ is generated in the component $g_{\rho \phi}$. Functions $\theta$, $\lambda$ and $\gamma$ are arbitrary, the latter describing the shape of the horizon. Boundary conditions (\ref{theBC}), apart from gathering the physically relevant solutions, yield finite and integrable charges. Other boundary conditions exist, which yield an additional super-translation current; however, the latter lead to nonintegrable charges. See also the interesting references \cite{Carlip1, Medved} for different criteria for selecting boundary conditions at the horizon. As we will see, our boundary conditions (\ref{theBC}) break Poincar\'e symmetry. 

The constant $\kappa$ corresponds to the black hole surface gravity. Our boundary conditions assume that $\kappa$ is a \emph{fixed constant without variation}, i.e., they describe the spectrum of black holes at fixed Hawking temperature $T={\kappa}/({2\pi})$. In the case of nonextremal BTZ black hole, this is given by
\begin{equation}\label{kappa3D}
\kappa = \frac{r_+^2 - r_-^2}{\ell^2 r_+},
\end{equation} 
where $r_+$ and $r_-$ are the outer and inner horizons.  
 
The asymptotic Killing vectors preserving the above asymptotic boundary conditions are
\begin{equation}\label{killing}
\begin{split}
\chi ^v &= T(\phi )+\mathcal{O}(\rho^3), \\
\chi ^{\rho } &= \frac{\theta}{2\gamma^2} T'(\phi) \rho^2+\mathcal{O}(\rho^3), \\
\chi ^{\phi } &= Y(\phi )-\frac{1 }{\gamma^2} T'(\phi) \rho + \frac{\lambda}{2\gamma^4} T'(\phi) \rho^2+ \mathcal{O}(\rho^3),
\end{split}
\end{equation}
where $T(\phi)$ and $Y(\phi)$ are arbitrary functions and the prime stands for the derivative with respect to $\phi$. Under such transformation, the arbitrary functions $\gamma(\phi)$ and $\theta(\phi)$ transform as
\begin{align}
\label{delta1}
\delta_\chi \theta = (\theta Y)'-2\kappa T',  \quad    \delta_\chi \gamma = (\gamma Y)'.
\end{align}
The asymptotic Killing vectors depend on fields defined on the metric. Accordingly, the algebra spanned by Lie brackets does not close. However, by introducing a modified version of Lie brackets \cite{Barnich3}
\begin{equation}
\label{modifed}
  [\chi_1,\chi_2]=\mathcal{L}_{\chi_1}\chi_2-\delta_{\chi_1} \chi_2+\delta_{\chi_2} \chi_1 ,
\end{equation}
one finds that the algebra of the asymptotic Killing vectors is given by
\begin{equation}
  [\chi(T_1,Y_1),\chi(T_2,Y_2)]=\chi(T_{12},Y_{12}),
\end{equation}
where
\begin{equation}
\begin{split}
T_{12}=Y_1 T_2'-Y_2T_1',\\
Y_{12}=Y_1 Y_2'-Y_2 Y_1'.  
\end{split}
\end{equation}
By defining Fourier modes, $T_n=\chi(e^{in\phi},0)$ and $Y_n=\chi(0,e^{in\phi})$ we find
\begin{equation}\label{algebra}
\begin{split}
i[ Y_m , Y_n ] &= (m-n) Y_{m+n}, \\  
i[ Y_m , T_n ] &=  - n T_{m+n},\\
i[ T_m , T_n ] &= 0.
\end{split}
\end{equation}
This is a semidirect sum of the Witt algebra generated by $Y_n$ with an Abelian current $T_n$.  The set of generators $Y_{-1}$, $Y_0$, $Y_{1}$ and $T_0$ form a $\mathfrak{sl}(2,\mathbb{R})\oplus\mathbb{R}$ subalgebra.

The $T_n$ generator is a supertranslation associated with the symmetry,
\begin{align}
v \to v+T(\phi),
\end{align}
already observed by Hawking \cite{Hawking} in four dimensions. In the current analysis, we have extended this symmetry by adding a vector field $Y_n$ which is responsible of generating superrotations 
\begin{align}
\phi \to \phi+Y(\phi),
\end{align}
on the circle of the horizon geometry.

Transformations \eqref{killing} have associated conserved charges at the horizon $\rho=0$. When considering three-dimensional Einstein gravity, these can be calculated in the covariant approach \cite{Barnich:2001jy}, yielding the charges
\begin{align}\label{Q}
Q(\chi) = \frac{1}{16\pi G}\int^{2 \pi}_0d\phi \left[ 2\kappa T(\phi) \gamma (\phi ) - Y(\phi ) \theta (\phi) \gamma(\phi) \right].
\end{align}
Their Poisson bracket algebra can be computed by noticing that, canonically, these charges generate the transformations \eqref{delta1}, i.e., $\{Q(\chi_1),Q(\chi_2)\}=\delta_{\chi_2} Q(\chi_1)$. In Fourier modes, $\mathcal{T}_n=Q{(T=e^{in\phi},Y=0)}$ and $\mathcal{Y}_n=Q{(T=0,Y=e^{in\phi})}$, the algebra spanned by $\mathcal{T}_n$ and $\mathcal{Y}_n$ is isomorphic to \eqref{algebra}, with no central extensions. 

It is worthwhile noticing that by defining the generator
\begin{align}\label{PT2}
\mathcal{P}_n=\sum_{k \in \mathbb{Z}} \mathcal{T}_k \mathcal{T}_{n-k}
\end{align}
the algebra spanned by $\mathcal{P}_n$ and $\mathcal{Y}_n$ is $\mathfrak{bms}_3$ \cite{Barnich1}
\begin{equation}
\begin{split}
i[ \mathcal{Y}_m , \mathcal{Y}_n ] &= (m-n) \mathcal{Y}_{m+n}, \\  
i[ \mathcal{Y}_m , \mathcal{P}_n ] &= (m - n) \mathcal{P}_{m+n},\\
i[ \mathcal{P}_m , \mathcal{P}_n ] &= 0.
\end{split}
\end{equation}
Therefore, although our asymptotic symmetries do not contain a Poincar\'e subgroup, the full BMS symmetry is recovered by means of the above Sugawara construction \cite{Grumiller}.

\subsubsection*{Exact solution}

Three-dimensional Einstein gravity in the presence of a negative cosmological term allows us to find an exact solution satisfying the above asymptotic boundary conditions, including BTZ black hole as a particular case. Its line element is \eqref{ds2}, where the functions read 
\begin{equation}\label{solexact}
\begin{split}
f&=-2\kappa \rho+\rho^2\left(\frac{\b ^2}{4 \gff^2}-\frac{1}{\ell^2}     \right)  ,\\
k&=1,\\
h&=\b \rho +\rho^2 \frac{\b}{4 \gff^2}\lambda(\phi),\\
R&=\gff +\rho \frac{\lambda(\phi)}{2 \gff},
\end{split}
\end{equation}
and where $\lambda$ is defined by
\begin{equation}
 \kappa \lambda (\phi)= \db-\frac{1}{2}\b^2+\frac{2}{\ell^2}\gff^2- \b \frac{\dgff}{ \gff}.
\end{equation}
$\b$ and $\gff$ are arbitrary functions, and $\ell$ stands for the AdS radius. The BTZ black hole is obtained by making the choice $\b={2 r_-}/{\ell}$ and $\gff=r_+$, while choosing $\kappa$ as \eqref{kappa3D}. In \cite{Lucietti}, a solution similar to \eqref{solexact} was presented, although with a different boundary condition on the function $R^{2}$.

It is interesting to study the special case $\kappa=0$ and $\theta=2 \gamma / \ell$.  For these values, the metric acquires the form
\begin{equation}\label{ds218}
ds^2= 2 dv d\rho + \frac{4}{\ell} \rho \gamma(\phi) dv d\phi+\gamma(\phi)^2d\phi^2,
\end{equation}
which has been found recently in the context of near horizon geometries of three-dimensional extremal black holes \cite{Compere:2015knw}. Note that the remaining symmetry algebra is just one copy of Virasoro.

When taking the flat limit $\ell \rightarrow \infty$, solution \eqref{solexact} also solves Einstein equations without cosmological constant.  After choosing $\kappa=-{J^2}/{2 r_H^3}$, its zero mode solution, i.e., $\theta={J}/{r_H}$ and $\gamma=r_H$,  corresponds to a flat cosmology with horizon radius $r_H$ \cite{Cornalba:2003kd}. 


The charges associated with solution \eqref{solexact} are given by \eqref{Q}. Evaluating for the case of the BTZ black hole, they read
\begin{equation}
\mathcal{T}_n= \frac{\kappa r_+}{4G} \delta_{n,0}, \quad \mathcal{Y}_n=- \frac{r_+ r_-}{4G \ell} \delta_{n,0}.
\end{equation} 
Hence, the charge associated with time translations $T_0 = \partial_v$ is the 
product of the black hole entropy $S=\pi r_+/(2G)$ and its temperature $T= \kappa/(2\pi)$. This means that the particular charge $\mathcal{T}_0 $, when varying the configuration 
space by fixing the temperature, corresponds to the entropy of the black hole. On the other hand, the charge associated with rotations along $Y_0=\partial_\phi$ coincides exactly with the angular momentum.

\subsection{Four-dimensional analysis}
\label{sec:3}
It is possible to extend the analysis of the first section to four dimensions. A suitable generalization of \eqref{ds2} is given by
\begin{equation}\label{ds2higher}
ds^2= f dv^2 + 2 k dv d\rho +2g_{vA}dv dx^A+g_{AB}dx^A dx^B ,
\end{equation}
where coordinates $x^A$ parameterize the induced surface at the horizon. The fall-off conditions on the fields as $\rho \to 0$ are 
\begin{equation}
\label{asym4D}
\begin{split}
f&= -2\kappa \rho + {\mathcal O}(\rho ^2), \\
k&=1+ {\mathcal O}(\rho ^2), \\
g_{vA}&= \rho \theta_A+{\mathcal O}(\rho^2 ), \\
g_{AB}&=\Omega\gamma_{AB}+\rho \lambda_{AB}+{\mathcal O}(\rho^2),
\end{split}
\end{equation}
while components $g_{\rho A}$ and $g_{\rho \rho}$ decay as ${\mathcal O}(\rho ^2)$ close to the horizon.
Here, $\theta_A$ and $\Omega$ are functions of the coordinates $x^A$, $\lambda^{AB}=\lambda^{AB}(v,x^A)$ and $\gamma_{AB}$ is chosen to be the metric of the two-sphere. It is convenient to use stereographic coordinates $x^A=(\zeta, \bar{\zeta})$ on $\gamma_{AB}$, in such a way that
\begin{align}
\gamma_{AB}dx^A dx^B=\frac{4}{(1+\zeta \bar{\zeta})^2} d\zeta d\bar{\zeta}.
\end{align}

The set of asymptotic conditions is preserved by the following vector fields:
\begin{equation}\label{killing2}
\begin{split}
\chi ^v &= T(\zeta, \bar{\zeta})+\mathcal{O}(\rho^3), \\
\chi ^{\rho} &= \frac{ \rho^2}{2\Omega}\theta_A\partial^A T+\mathcal{O}(\rho^3), \\
\chi ^A &= Y^A-\frac{\rho}{\Omega} \partial^A T+\frac{\rho^2}{2\Omega^2} \lambda^{AB} \partial_B T +\mathcal{O}(\rho^3),
\end{split}
\end{equation}
where we have used $\gamma^{AB}$ to raise indices and $Y^A$ is a function of $x^A$ only, i.e., $Y^{\zeta}=Y(\zeta)$ and $Y^{\bar{\zeta}}=\bar{Y}(\bar{\zeta})$.
Under these transformations, the fields transform as
\begin{equation}
\begin{split}
\label{delta}
  \delta_\chi \theta_A &= Y^B \partial_B \theta_A+\partial_A Y^B  \theta_B-2\kappa \partial_A T,\\ 
  \delta_\chi \Omega &=\nabla_B(Y^B \Omega),
\end{split}
\end{equation}
$\nabla$ standing for the covariant derivative on $\gamma_{AB}$. 

Under modified Lie brackets \eqref{modifed}, transformations \eqref{killing2} satisfy
\begin{equation}
  [\chi(T_1,Y^A_1),\chi(T_2,Y^A_2)]=\chi(T_{12},Y^A_{12}),
\end{equation}
where
\begin{equation}
\begin{split}
T_{12}&=Y_1^A \partial_A T_2-Y_2^A \partial_AT_1,\\
Y^A_{12}&=Y_1^B \partial_B Y_2^A-Y_2^B \partial_B Y_1^A.  
\end{split}
\end{equation}

Notice that the transformations generated by $Y^A$, in general, are not globally well defined on the two-sphere. The only invertible transformations are those spanning the global conformal group, which is isomorphic to the proper, orthochronous Lorentz group. However, if we focus only on the local properties, all functions are allowed. This was first proposed in \cite{Barnich2,Barnich3} in the context of asymptotically flat spacetimes. 

By expanding in Laurent modes
\begin{equation}
\begin{split}
 T_{(n,m)}&=\chi(\zeta^n \bar\zeta^m,0,0), \\ 
 Y_n&=\chi(0,-\zeta^{n+1},0), \\  
 \bar{Y}_n&=\chi(0,0,-\bar {\zeta}^{n+1}),
 \end{split}
\end{equation}
the nonvanishing commutation relations read
\begin{equation}
\label{alg4D}
\begin{split}
  &[Y_n,Y_m]=(n-m)Y_{n+m},\\
  &[\bar{Y}_n,\bar{Y}_m]=(n-m)\bar{Y}_{n+m},\\
  &[Y_k,T_{(n,m)}]=-n T_{(n+k,m)},\\
  &[\bar{Y}_k,T_{(n,m)}]=-m T_{(n,m+k)}.
  \end{split}
\end{equation}
The exact isometry algebra corresponds to $\mathfrak{sl}(2,\mathbb{C})\oplus \mathbb{R}$ whose elements correspond to the globally well-defined transformations on the sphere plus $T_{(0,0)}$.

Conserved charges at the horizon turn out to be given by
\begin{align}\label{Q4D}
Q{(T,Y^A)} = \frac{1}{16\pi G}\int d\zeta d\bar{\zeta} \sqrt{\gamma}\ \Omega \left[ 2\kappa T  - Y^A \theta_A \right].
\end{align}
They close under Poisson bracket 
\begin{align}\label{Q2}
\{Q(T_1,Y^A_1),Q(T_2,Y^A_2)\} =Q(T_{12},Y^A_{12}).
\end{align}
By defining $\mathcal{T}_{(m,n)}=Q(\zeta^n \bar\zeta^m,0,0)$, $\mathcal{Y}_n=Q(0,-{\zeta}^{n+1},0)$ and $\mathcal{\bar{Y}}_{n}=Q(0,0,-\bar{\zeta}^{n+1})$, we find that these quantities satisfy the same algebra \eqref{alg4D}.

We can perform the Sugawara construction as we did in the previous section. Defining 
\begin{eqnarray}
\mathcal{P}_{(n,l)}=\sum_{m\in\mathbb{Z}} \sum_{t\in\mathbb{Z}} \mathcal{T}_{(m,t)}\mathcal{T}_{(n-m,l-t)} 
\end{eqnarray}

 and using \eqref{alg4D}, one finds
\begin{eqnarray}
\label{algwithP}
\begin{split}
[\mathcal{P}_{(n,l)},\mathcal{Y}_m]&=(n-m)\mathcal{P}_{(n+m,l)} \\
[\mathcal{P}_{(n,l)},\bar{\mathcal{Y}}_m]&=(l-m)\mathcal{P}_{(n,l+m)}.
\end{split}
\end{eqnarray}
Although this is reminiscent of $\mathfrak{bms}_4$, notice that this is not exactly the same algebra as that found in \cite{Barnich2,Barnich3}.

Finally, let us note that Kerr black hole fits in our boundary conditions \eqref{asym4D}. An explicit construction of this solution in term of Gaussian normal coordinates can be found in \cite{Booth:2012xm}. One can verify that
\begin{equation}
\mathcal{T}_{(0,0)}= \frac{\kappa}{2\pi}\frac{ \mathcal{A}}{4G},\quad \mathcal{Y}_{0}=\frac{iaM}{2},\quad \mathcal{\bar{Y}}_{0}=-\frac{iaM}{2},
\end{equation}
where $\mathcal{A}$ is the area of the horizon, while $M$ and $a$ are the usual parameters of the Kerr solution. That is, the zero mode of the supertranslation is the product of the black hole entropy with its temperature. Since our boundary conditions are defined by fixing $\kappa$, we can associate this charge with Wald entropy. On the other hand, the charge  $Q(0,\partial_\phi)=-i(\mathcal{Y}_0-\mathcal{\bar{Y}}_0)=aM$ is the angular momentum.

In the case where $m$ and $n$ are different from zero, $\mathcal{Y}_{n}$, $\mathcal{\bar{Y}}_{n}$ and $\mathcal{T}_{(m,n)}$ with  $m \neq n$ vanish. In contrast, charges $\mathcal{T}_{(m,m)}$ with $m\neq0$ diverge. This phenomenon was first noticed in \cite{Barnich4} and has been explained in reference \cite{Barnich6}. Let us explain the origin of this divergence for the case of Schwarzschild black hole. In this case, the supertranslation charge reads
\begin{equation}
\mathcal{T}_{(m,n)}=\frac{\kappa r_+^2}{4G} \delta_{m,n} I(m) ,
\end{equation}
where  $I(m)=\int^{\pi}_0 d \theta \sin(\theta) \cot^{2m}(\theta/2)$ is divergent for $m\neq0$, with the divergence comes from the poles of the sphere. 
If instead of Laurent modes, the supertranslation $T(\zeta,\bar{\zeta})$ is expanded in spherical harmonics, the charges can be seen to vanish.
\vspace{3mm}
\section{Discussion}
\label{sec:discussion}
We have shown that the near horizon geometry of nonextremal black holes exhibits an infinite dimensional extension of supertranslation algebra, which in particular contains superrotations. This phenomenon is similar to what happens in the asymptotically flat spacetimes at null infinity, although the algebra obtained differs from the standard extended BMS. We have explicitly worked out the cases of three-dimensional and four-dimensional stationary black holes, for which the zero modes of the charges associated to the infinite-dimensional symmetries were shown to exactly reproduce the entropy and the angular momentum of the solutions. 

In the three-dimensional case, we have presented a family of explicit solutions that obey the proposed boundary conditions at the horizon and, therefore, realize the infinite-dimensional symmetry generated by the semidirect sum of Virasoro algebra and supertranslations. Although this family of solutions represent locally AdS$_3$ spacetimes, they do not satisfy the standard Brown-Henneaux asymptotic conditions at $\rho \to \infty$, as we are imposing boundary conditions at the horizon $\rho \to 0$. In \cite{CompereStrominger}, a set of asymptotically AdS$_3$ boundary conditions were found whose associated charges yield a centrally extended version of algebra (\ref{algebra}). It would be interesting to study the relation between such boundary conditions and (\ref{theBC}); in particular, to clarify the precise connection between the family of solutions (\ref{solexact}) and those presented in \cite{CompereStrominger}. The latter also includes the BTZ black hole as a particular example; however, in contrast to (\ref{theBC}), which fixes the black hole surface gravity $\kappa $, the boundary conditions considered in \cite{CompereStrominger} are defined by fixing the value of $\Delta = M\ell +J$.  

Another question is whether it is possible to modify our boundary conditions in such a way of getting nonvanishing central extensions. In this regard, it is worthwhile mentioning that the boundary conditions we have considered allow for exponentially decaying modes $e^{-\kappa v} X(\phi)$ which yield an extra infinite-dimensional symmetry also associated to an extension of supertranslations. On the other hand, an important point to address is the study of the extremal limit, for which the boundary conditions at the horizon need to be reconsidered since the leading term in $g_{vv}$ vanishes. Finally, it would be worthwhile investigating whether this infinite dimension symmetry can have applications to memory effects in black hole physics.

To conclude, let us mention that the idea of investigating the symmetries of the horizon has been considered for a long time by different authors; see for instance \cite{Carlip2}. Infinite-dimensional symmetries were discussed in a similar context in \cite{Koga,Hotta:2002mq,Hotta:2000gx}. It would be interesting to study the connection between those works and ours.

\[ \]
The authors thank Glenn Barnich, Geoffrey Comp\`ere and C\'edric Troessaert for useful discussions. L.D.~is a
FRIA research fellow of the FNRS Belgium. The work of G.G. has been supported by CONICET. The work of H.G. is  supported in part by the Fund for Scientific
Research-FNRS (Belgium), by IISN-Belgium and by ``Communaut\'e fran\c{c}aise de
Belgique - Actions de Recherche Concert\'ees''. The work of M.P. is funded by FONDECYT/Chile grants 7912010045 and 11130083.
\[ \]

\providecommand{\href}[2]{#2}\begingroup\raggedright\endgroup
\end{document}